\documentclass[twocolumn,showpacs,preprintnumbers,amsmath,amssymb,aps]{revtex4}

\usepackage{graphicx}   
\usepackage{dcolumn}    
\usepackage{bm}         

\begin{document}


\title{Electron cyclotron mass in undoped CdTe/CdMnTe quantum wells}

\author{A.A.~Dremin$^{1,2}$}
\author{D.R.~Yakovlev$^{1,3}$}
\author{A.A.~Sirenko$^{4}$}
\author{S.I.~Gubarev$^{2}$}
\author{O.P.~Shabelsky$^{2}$}
\author{A.~Waag$^{5}$}
\author{M.~Bayer$^{1}$}

\affiliation{ $^{1}$Experimentelle Physik II, University of
Dortmund, D-44227 Dortmund, Germany \\
$^{2}$Institute of Solid State Physics, Chernogolovka,
Moscow district, 142432, Russia \\
$^{3}$A.F.Ioffe Physico-Technical Institute, Russian
Academy of Sciences, 194021 St. Petersburg, Russia \\
$^{4}$New Jersey Institute of Technology, Newark, NJ
07102, USA \\
$^{5}$Institute of Semiconductor Technology, Braunschweig
Technical University, 38106 Braunschweig, Germany
}

\date{\today}

\begin{abstract}
Optically detected cyclotron resonance of two-dimensional
electrons has been studied in nominally undoped CdTe/(Cd,Mn)Te
quantum wells. The enhancement of carrier quantum confinement
results in an increase of the electron cyclotron mass from
0.099$m_0 $ to 0.112$m_0 $ with well width decreasing from 30 down
to 3.6~nm. Model calculations of the electron effective mass have
been performed for this material system and good agreement with
experimental data is achieved for an electron-phonon coupling
constant $\alpha $=0.32.
\end{abstract}


\maketitle


\section{Introduction}

The effective masses of carriers (electrons and holes) are among
the basic parameters for semiconductors and semiconductor
heterostructures. Nowadays exhaustive information is available for
heterostructures based on III-V semiconductors, $e.g.$
GaAs/(Al,Ga)As heterosystems. However, only limited experimental
data have been reported so far for the II-VI family of
semiconductor heterostructures. Among them are the structures
based on CdTe, which are rather popular for optical studies. One
of the attractions to this material is the possibility to
introduce magnetic Mn-ions in the cation sublattice. The strong
exchange interaction of free carriers with localized spins of
magnetic ions gives rise to giant magneto-optical effects, $e.g$.
the giant Zeeman splitting of the band states, giant Faraday
rotation, \textit{etc}. \cite{Furd88}. (Cd,Mn)Te, (Cd,Mg)Te and
(Cd,Zn)Te are among the barrier materials to confine carriers in
CdTe quantum wells. In this paper we study experimentally the
dependence of the electron effective mass on quantum well (QW)
width for CdTe/(Cd,Mn)Te heterostructures.

The cyclotron resonance (CR) technique is widely used for
evaluation of the fundamental parameters of heterostructures,
including the carriers effective masses. It has been recently
applied to modulation-doped CdTe/(Cd,Mg)Te QWs and the electron
effective mass has been measured for these QWs with widths varied
between 7.5 and 30 nm \cite{Karc02,Iman98} each, with a 2D
electron gas density of 4$\times $10$^{11}$ cm$^{ - 2}$. One of
the impediments for the conventional cyclotron resonance technique
is that the carrier density has to be large enough to produce a
noticeable change in the absorption of microwave or far-infrared
(FIR) radiation. This limitation does not allow to measure carrier
effective masses in undoped systems. It has been overcome by
invention of the Optically Detected of Cyclotron Resonance (ODCR,
or ODR) technique (see \cite{Godl94} and references therein).

The ODR technique is based on variation of the optical properties,
such as the photoluminescence intensity under absorption of
microwave or FIR radiation by free carriers. It was proved to be
extremely sensitive and has been successfully used to measure the
effective masses of electrons and holes in bulk GaAs, InP, CdTe
\cite{Rome80, Moll92, Mich94}, and SiC \cite{Chen97}. It was also
developed to study 2D electron states in GaAs/(Al,Ga)As
heterostructures \cite{Guba91, Ahme92, Warb92} and internal
transitions of neutral and charged magnetoexcitons \cite{Nick00,
Nick02, Mein03}. Another advantage of the ODR technique is related
to its spectral selectivity, which allows for selecting the signal
from different quantum wells grown in the same structure by
analyzing the corresponding photoluminescence emission lines.
Therefore, the ODR technique is very well suited for measurements
of the electron effective masses in undoped CdTe-based QWs of
different widths.

\section{Experiment}

We have studied a CdTe/Cd$_{0.86}$Mn$_{0.14}$Te quantum
heterostructure grown by molecular beam epitaxy on an
(100)-oriented CdTe substrate. The structure contains four
consequently grown CdTe wells with width $L_Z = $30, 9, 3.6 and
1.2~nm separated by 50-nm-thick Cd$_{0.86}$Mn$_{0.14}$Te barriers
from each other. Typical photoluminescence (PL) and reflectivity
spectra of such structures can be found in \cite{Waag91, Ivch92,
Yako95}.

\begin{figure*}[tbp]
\includegraphics[width=.66\textwidth]{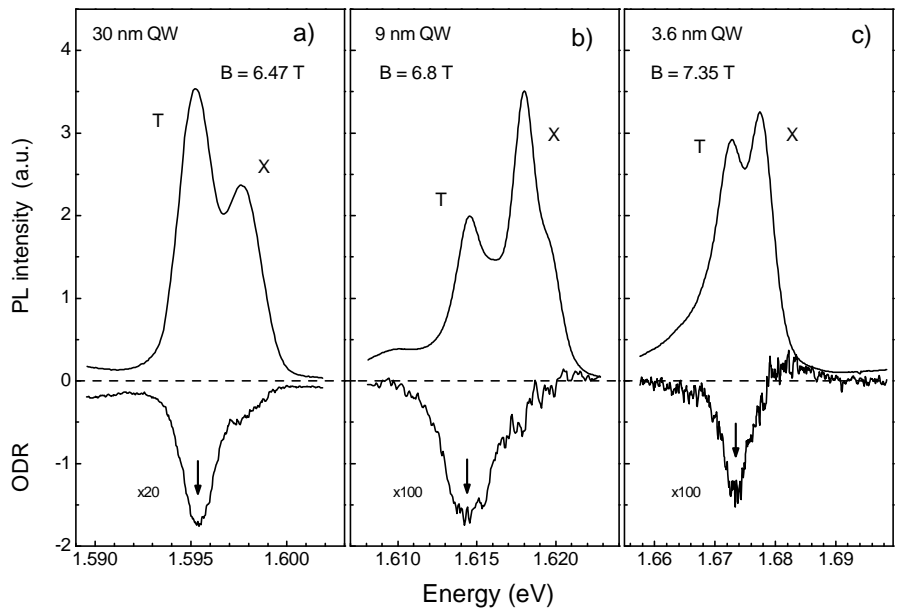}
\caption{Photoluminescence and ODR signal spectra measured for
CdTe/Cd$_{0.86}$Mn$_{0.14}$Te QWs with widths equal to: (a) 30~nm,
(b) 9~nm, and (c) 3.6~nm. Spectra are measured at magnetic fields
for which the FIR radiation induces the maximal changes, i.e.
under conditions of cyclotron resonance for electrons. $T$=4.2K.
The photon energies corresponding to the maximum of the ODR signal
are marked with arrows. Note that the magnetic fields scans of the
cyclotron resonances shown in Fig.~\ref{fig3} were detected at
these energies.} \label{fig1}
\end{figure*}

Experiments were carried out at a temperature of $T = 4.2$~K in a
He exchange cryostat in magnetic fields up to $B = 8.3$ T. The
sample was mounted on a rotating platform, which enables ODR
measurements in tilted magnetic fields in order to check the
two-dimensional character of the studied resonances. Most
experimental data were collected in magnetic fields oriented
parallel to the structure growth axis ($\theta = 0^{\circ})$ with
the cyclotron motion of electrons in the plane of the quantum
wells. Photoexcitation of the samples by a HeNe laser and
collection of the luminescence signal was provided via optical
fibers. The spot of the HeNe laser beam ($\lambda = 6328$~{\AA},
power up to 20~mW) was overlapped by the spot of a CO$_2 $ pumped
FIR laser. Far infrared (FIR) radiation ($\lambda _{FIR} =
163$~$\mu $m, $E_{FIR} $=7.6~meV) with a power up to 15~mW was
guided into the cryostat via a stainless steel pipe and focused on
the sample by a Teflon lens. Photoluminescence (PL) signal was
analyzed with a 0.6~m grating spectrometer equipped with a cooled
photomultiplier.

The FIR laser beam was mechanically chopped. The influence of the
FIR radiation on the PL spectra was synchronously detected by a
lock-in amplifier at various magnetic fields. The ODR signal was
normalized to the PL intensity $I(B)$ measured at the same
wavelength. This procedure allowed us to correct the shape of the
resonance profile by accounting for the PL intensity variations
with increasing magnetic field.

\section{Results and discussion}

Photoluminescence spectra for three
CdTe/Cd$_{0.86}$Mn$_{0.14}$Te\textbf{ }QWs are shown in
Fig.~\ref{fig1}. The emission spectra of all three QWs consist of
two strong lines corresponding to excitons (X) localized at
well-width fluctuations and to charged exciton complexes, $i.e.$
trions (T) consisting of two electrons and one hole \cite{Jeuk02}.
Their formation requires excess of electrons over holes in QWs.
Such excess is typical for unintentionally doped CdTe QWs, due to
carrier diffusion from the barrier materials with residual n-type
doping. The energy difference between the exciton and trion lines
varies from 2.5 to 5~meV and increases in narrow wells. It
corresponds to the trion binding energy, which is about an order
of magnitude smaller than the binding energy of the quasi-2D
excitons.

\begin{figure}[tbp]
\includegraphics[width=.45\textwidth]{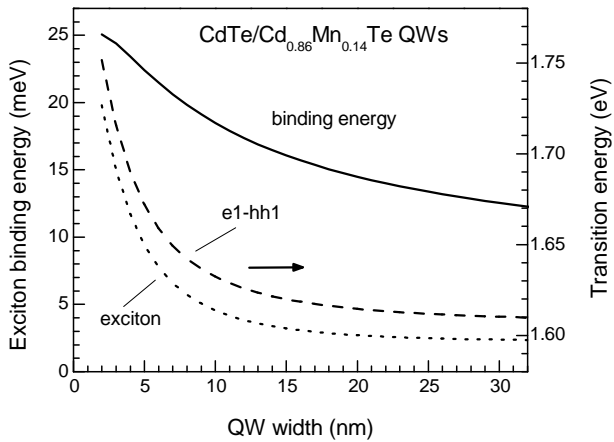}
\caption{Exciton energy, exciton binding energy and energy of
optical transition between the lowest levels of confined electrons
and holes (e1-hh1) calculated for CdTe/Cd$_{0.86}$Mn$_{0.14}$Te
QWs as function of the QW width.} \label{fig2}
\end{figure}

One can also see in Fig.~\ref{fig1} that the exciton emission line
shifts from 1.597 up to 1.678~eV for QW width varied from 30 down
to 3.6~nm due to the carrier quantum confinement. We use the
energy position of the exciton emission to evaluate the QW width
for the studied samples. Model calculations for the exciton PL
transition energy and the exciton binding energy have been
performed using the procedure described in \cite{Asta02} with the
following parameters for our material system: the band gap offset
between the well and barrier materials is 223~meV, it is divided
in a ratio of 70/30 between the conduction and valence bands; the
dielectric constant $\varepsilon $=10; the in-plane heavy-hole
mass was taken as $m_{hh,\parallel } $=0.37$m_0 $; the heavy-hole
mass along the growth axis, $i.e$. perpendicular to the QW plane,
is $m_{hh, \bot } $=0.48$m_0 $. We have taken into account eight
confined electron levels and ten confined hole levels. The results
of our calculations are presented in Fig.~\ref{fig2}, from which
the widths of the quantum wells have been deduced by comparing the
experimental exciton PL transition energies with the calculated
dependence.

We turn now from the sample characterization to the results of the
optically detected resonance. ODR signal could be reliably
detected in the QWs with widths $L_Z = $30, 9 and 3.6~nm. We have
found no influence of FIR on the emission from the narrowest QW
with $L_Z = $1.2~nm. Most probably the changes are below the
sensitivity level of our setup.

\begin{figure}[tbp]
\includegraphics[width=.4\textwidth]{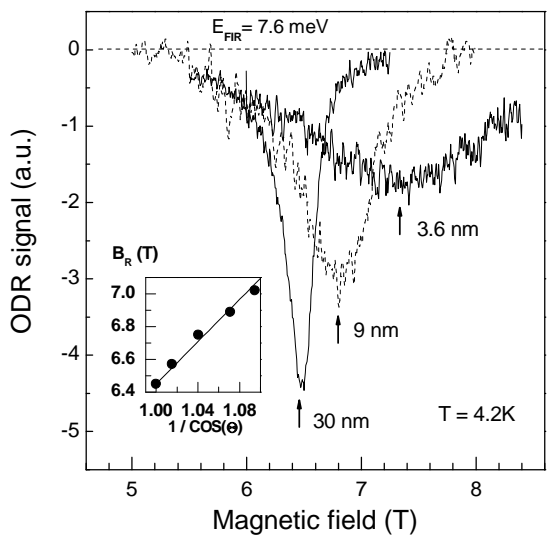}
\caption{Magnetic field dependence of ODR signal at $E_{FIR} =
7.6$~meV measured in CdTe/Cd$_{0.86}$Mn$_{0.14}$Te QWs. The shift
of the resonance field is given in the insert as function of the
tilt angle $\theta $ for the 30-nm-wide QW (circles -- experiment
values; line -- fit with 6.45/$\cos (\theta ))$.}\label{fig3}
\end{figure}

The ODR signal intensity plotted as a function of magnetic field
clearly demonstrates a resonance behavior (Fig.~\ref{fig3}). We
have checked that both the resonance field and the shape of the
resonance profile are insensitive to the PL detection energy. The
ODR signal was recorded at fixed detection energies shown by the
arrows in Fig.~\ref{fig1}. The ODR signal was normalized to the PL
intensity measured at the same detection energy. For all QWs the
PL intensity was decreasing by about 30{\%} with increasing
magnetic field from 5.5~T to 8~T. This change has very small
influence on the shape of the resonance profile and requires
correction of the resonance field by less than 0.01~T.

There are three characteristics of the resonance curves to be
analyzed: (i) the resonance magnetic field $B_R $, which is
directly linked to the value of the electron effective mass; (ii)
the resonance full width at the half maximum (FWHM), which is
inversely proportional to the electron scattering rate and
contains information on the electron mobility, and (iii) the
resonance amplitude, which is controlled by the mechanisms
responsible for the ODR signal. Before proceeding with discussion
of the resonance parameters we shall proof that the observed
features originate from the QWs. In order to check the
two-dimensional character of electrons responsible for the ODR
signal we have carried out the same measurements in tilted
magnetic fields. The pronounced shift of the CR resonance toward
higher magnetic fields was found to be proportional to 1 /
cos$\theta $ where $\theta $ (being varied from 0\r{ } to 24\r{ })
is the angle between the magnetic field direction and the
structure growth axis. The insert in Fig.~\ref{fig3} illustrates
this observation for the 30~nm QW proving that the electrons have
quantum confined character.

In the widest QW with 30~nm width, the resonance FWHM is 0.33~T.
This corresponds to a momentum relaxation time of 2.4~ps, and to
an electron mobility of 3.0$\times $10$^{4}$~cm$^{2}$/(V$ \cdot
$s). A decrease of the well width is accompanied by a strong
broadening of the resonances up to 0.74~T and 1.9~T for 9~nm and
3.6~nm wells, respectively. The corresponding electron mobilities
are 1.4$\times $10$^{4}$ and 0.5$\times $10$^{4 }$~cm$^{2}$/(V$
\cdot $s). Localization of electrons on QW width fluctuations is
known to be the dominating mechanism for resonance broadening in
low dimensional structures. Its contribution increases in narrow
QWs causing the decrease of the carrier mobility. It goes in line
with the increasing width of the photoluminescence emission
spectra from 1.7~meV to 3.7~meV for 30~nm and 3.6~nm wells,
respectively (see Fig.~\ref{fig1}).

The modulation spectra (ODR signal) recorded at the resonance
magnetic field are shown in the lower panels of Fig.~\ref{fig1}.
In the 30~nm QW the FIR radiation results in a decrease of the PL
signal by approximately 2.5{\%} for the trion line and a
significantly smaller decrease for the exciton line. The ODR
signal decreases in narrower QWs, it is about 0.9{\%} for the 9~nm
QW and only 0.5{\%} for the 3.6~nm QW. This observation can be
attributed to enhanced electron localization and the related
decrease of the electron mobility in narrow QWs.

The dominating mechanism of the PL intensity modulation under FIR
radiation is related to the specifics of the trion complexes in
the studied structures. As one can see in Fig.~\ref{fig1}, the
strongest ODR signal has been observed in the maximum of the trion
emission line and only weak modulations are seen for the exciton
line. This is expected since in QWs with a very diluted electron
gas the trion emission is much more sensitive to the temperature
of the electron gas \cite{Jeuk02, Hu98} then the exciton emission.
This is due to the fact that for the trion formation one of the
electrons is captured from the electron gas and, hence, the
probability of the trion formation is very sensitive to the
electron gas temperature. Heating of the electrons under cyclotron
resonance conditions decreases the probability for trion
formation, which causes a decrease of the trion emission
intensity.

\begin{figure}[tbp]
\includegraphics[width=.4\textwidth]{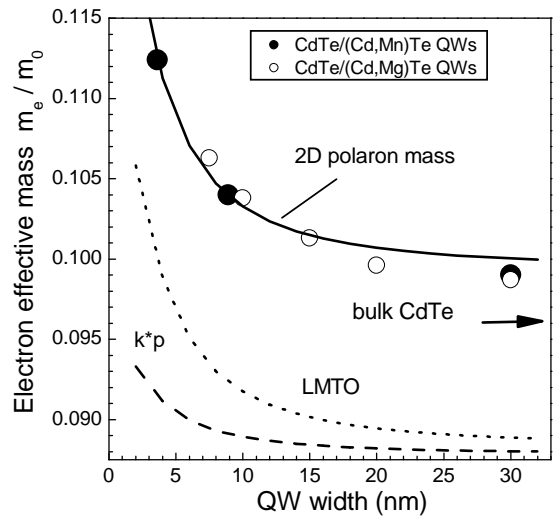}
\caption{Electron cyclotron mass versus QW width for CdTe-based
quantum wells. Experimental data of this work for
CdTe/Cd$_{0.86}$Mn$_{0.14}$Te QWs are given by closed circles.
Open circles show the data for the modulation-doped
CdTe/Cd$_{0.88}$Mg$_{0.12}$Te QWs with electron density of
4$\times $10$^{11}$~cm$^{ -2}$ \cite{Karc02}. The bare electron
mass calculated without polaron correction in parabolic-band
approximation is shown by the dashed line and that with accounting
for conduction band non-parabolicity is given by the dotted line.
The solid line shows the calculation after including the
electron-phonon interaction with $\alpha = 0.32$.} \label{fig4}
\end{figure}

Measurement of the resonance magnetic field $B_R $, where the FIR
energy coincides with the cyclotron energy of electrons, allows
evaluation of the electron effective mass. Fitting the resonances
shown in Fig.~\ref{fig3} by a Lorenzian function we obtained $B_R
$ = 6.45, 6.75 and 7.25~T for QWs with $L_Z $ = 30, 9 and 3.6~nm,
respectively. The electron cyclotron mass $m_e $ was evaluated
from $B_R $ values using ${m_e } \mathord{\left/ {\vphantom {{m_e
} {m_0 = 0.0152\,\,B_R \,\left[ \mbox{T} \right]}}} \right.
\kern-\nulldelimiterspace} {m_0 = 0.0152\,\,B_R \,\left[ \mbox{T}
\right]}$, which is derived from $m_e = {e\hbar B_R }
\mathord{\left/ {\vphantom {{e\hbar B_R } {E_{FIR} }}} \right.
\kern-\nulldelimiterspace} {E_{FIR} }$ for $E_{FIR} $= 7.6 meV. We
found that $m_e $ increases with decreasing well width: $m_e =
0.099\,m_0 $, $0.104\,m_0 $ and $0.112\,m_0 $ for $L_Z = $~30, 9
and 3.6~nm. These data are shown by solid circles in
Fig.~\ref{fig4}. The arrow in the figure marks the electron
effective mass of $0.096\,m_0 $ measured for bulk CdTe
\cite{Land82}. The open circles are experimental data for
CdTe/(Cd,Mg)Te QWs with non-magnetic barriers taken from
\cite{Karc02}. These results coincide well with our experimental
data. For the studied relatively wide QWs the dominating part of
the electron wave function is concentrated in the CdTe wells and
is not much dependent on the differences in barrier materials.
Also, the (Cd,Mn)Te and (Cd,Mg)Te alloys are pretty similar in
their properties as these barrier materials provide efficient
confinement of both electrons and holes. One may expect that in
doped CdTe/(Cd,Mg)Te QWs $m_e $ will be larger due to
non-parabolicity of the conduction band at finite $k$-values given
by the Fermi level. However, an estimation of this effect for the
electron density of 4$\times $10$^{11}$ cm$^{ - 2}$ gives an $m_e
$ increase by 1.2{\%} only \cite{Karc02}, which does not exceed
the error bar for our experimental data. Comparing the data for
the two systems with different barrier materials we can conclude
that the electron confinement and the respective increase of the
quantum confinement energy is the dominating factor in the $m_e $
dependence on the QW width.

For deeper insight into the electron mass behaviour we have
compared the experimental data with results of model calculations.
We will follow the commonly used routine in which the bare
effective mass $m_b $ is calculated and the polaron correction to
the effective mass is fitted with the coupling constant of the
electron-phonon interaction $\alpha $ used as a free parameter

\begin{equation}
\label{eq1} m_e = m_b \,\,(1 + \pi \alpha / 8).
\end{equation}

The polaron contribution to the measured carrier effective masses
originates from the cloud of optical phonons that ``accompany''
the electron and make the measured mass ``heavier''. For example,
in bulk CdTe the calculated bare electron effective mass is
$0.088\,m_0 $ and the measured polaron mass is $0.096\,m_0 $.

A simple analytical approach to calculation of $m_b $ is provided
in the frame of 3-band \textbf{k$ \cdot $p} technique

\begin{equation}
\label{eq2} \frac{m_0 }{m_b } = \left( {1 + 2F} \right) +
\frac{E_P \left( {E_g + {2\Delta _{SO} } \mathord{\left/
{\vphantom {{2\Delta _{SO} } 3}} \right.
\kern-\nulldelimiterspace} 3} \right)}{E_g \left( {E_g + \Delta
_{SO} } \right)},
\end{equation}
where $E_P $= 21.0~eV is the interband matrix element, $\Delta
_{so}$ = 0.93~eV is the spin-orbit splitting, $E_{g}$ = 1.59~eV is
the fundamental band gap of CdTe, and $F$ = -0.6 is a parameter
accounting for contribution of high energy bands \cite{Card88,
Will95, Sire97}. One can see from Eq.(\ref{eq2}) that the
effective mass $m_b $ increases for larger band gap values. Exact
calculation of $m_b $ in QW structures requires laborious
numerical procedures, which include the quantum confinement
energy, conduction band non-parabolicity, and penetration of the
electron wave function into barriers (see $e.g$. \cite{Smit90}).
However, the 3-band \textbf{k$ \cdot $p} approach gives fairly
well suited values when $E_g $ in Eq.(\ref{eq2}) is replaced by
the energy separation between the lowest electron and hole
subbands. In Fig.~\ref{fig4} these calculations for CdTe-based QWs
in parabolic band approximation are shown by the dashed line. In
order to take into account the conduction band non-parabolicity in
CdTe, we performed more elaborate calculations based on density
functional theory in local density approximation using the linear
muffin-tin-orbital (LMTO) approach \cite{Savr96}. Calculated
dependence for the bare electron mass (the dotted line in
Fig.~\ref{fig4}) demonstrates the main experimental trend of
increasing mass values with narrowing QW width.

Very good agreement with experimental data has been achieved by
correcting the bare mass with the polaron effect using
Eq.(\ref{eq1}). The best fit shown by the solid line has been
achieved for $\alpha $ = 0.32. This value is in good agreement
with the literature data for the electron-phonon interaction
constant reported for CdTe and CdTe/(Cd,Mn)Te QWs (see
\cite{Nich94} and references therein).

To conclude, the optically-detected resonance technique has been
used to study electron cyclotron resonance in nominally undoped
CdTe/(Cd,Mn)Te QWs. Pronounced modulation of the luminescence
intensity has been found for the charged exciton emission, when
the residual electrons are resonantly heated by FIR radiation. The
evaluated electron effective masses increase with narrowing
quantum well width and are in good agreement with data for CdTe
QWs confined by (Cd,Mg)Te barriers. Good quantitative agreement
with model calculations is achieved with an electron-phonon
interaction constant $\alpha $ equal to 0.32.

\begin{acknowledgments}
This work was supported by the QuIST program of DARPA, the
'Forschugsband Mikro- und Nanostrukturen' of the University of
Dortmund, the BmBF project 'nanoquit' and the Russian Foundation
for Basic Research.
\end{acknowledgments}

\end{document}